\documentclass[twocolumn,showpacs,preprintnumbers,amsmath,amssymb]{revtex4}

\usepackage{graphicx}
\usepackage{dcolumn}
\usepackage{bm}
\usepackage{amssymb}

\def\address{\@ifstar{\address@star}%
  {\@ifnextchar[{\address@optarg}{\address@noptarg}}}

\begin{document}

\author{S.N.~Gninenko\footnote{ Sergei.Gninenko\char 64 cern.ch}}

\affiliation{Institute for Nuclear Research of the Russian Academy of Sciences, Moscow 117312}


\title{ Limit on the electric charge-nonconserving $\mu^+ \to invisible$  decay}

\date{\today}

\begin{abstract}
The  first limit on the branching ratio of the electric charge-nonconserving
invisible muon decay 
$Br(\mu^+ \to invisible) < 5.2 \times 10^{-3}$ is obtained from the
 recently reported results on 
new determination of the Fermi constant from  muon decays. 
The results of a feasibility study of a new proposed experiment 
 for a sensitive search for this decay mode
at the level of a few parts in $10^{11}$  are presented. 
 Constrains on the $\tau \to invisible$ decay rate  
are discussed.  These leptonic charge-nonconserving processes may 
hold in four-dimensional world in models with infinite extra dimensions, thus 
making their searches  complementary to collider experiments probing 
new physics. 
\end{abstract}
\maketitle

\section{Introduction}

The law of the electric charge conservation, related to the existence of the 
unbroken local U(1) 
gauge symmetry of the Standard Model (SM), is thought to be an absolute 
conservation law. However,  ``...the basic principles of theoretical physics 
cannot be accepted {\it a priori}, no matter how convincing they may seem, 
but rather must be justified on the basis of relevant experiments'' \cite{fg}.
Thus, new experimental tests of old and fully 
established laws of physics are of importance 
as they could provide new unexpected results \cite{oz}-\cite{all}.   

The quantization of the electric charge has also been established 
with a high level of accuracy, while   the question of why the electric 
charge is quantized seems has yet unknown answer.
Going beyond the SM, it is    
possible to have particles with a
 small fractional (milli) charge, e.g. by introducing 
additional U'(1)  symmetry of a hidden-sector \cite{okun, bob}. 
These considerations have stimulated new theoretical works
and experimental tests reported in Ref.\cite{milli}.\\

Not long ago, it has been pointed out that 
in some models with infinite extra dimensions,  describing our 
world as a brane embedded in higher dimensional space \cite{rs,rasu},
 particles initially 
located on our brane may leave the brane and disappear into extra dimensions
\cite{dr,drt1}. The experimental signature of this 
effect is the disappearance of the particles in our world, 
i.e. the $particle \to invisible$ decay.
These transitions have been found to be generic in a class of models
of localization of particles on a brane. 
The localization becomes 
incomplete if particles get small masses and they could tunnel from the brane 
into extra dimensions \cite{drt2,rub}.

An example of this process  for a neutral system, 
orthopositronium ($o-Ps$), a triplet bound state of an electron 
and positron, has been considered in Ref.\cite{gkr}. It has been shown 
that the effect  may occur at a rate within
 two orders of magnitude of the present best experimental limit
on the  branching ratio of the $o-Ps \to invisible$ decay 
$Br(o-Ps \to invisible) < 4.3 \times 10^{-7}$ (90\% C.L.) 
from the recent ETH-INR experiment \cite{bader,pc}. 

\begin{figure}[htb!]
\includegraphics[width=0.4\textwidth]{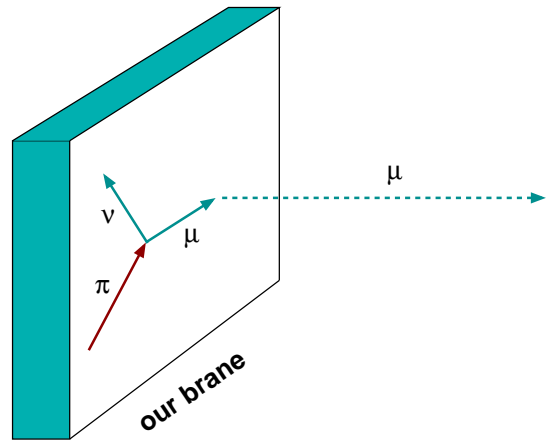}
\caption{\label{fig:brane} Schematic illustration of the disappearance of muon
 produced in the $\pi \to \mu + \nu$ decay: neutrino propagates along the 
our brane while the muon escapes into extra dimensions.} 
\end{figure}

If, however, particles that leave our brane are electrically charged, their 
disappearance into extra dimensions would result in the nonconservation of 
electric charge seen by  our four-dimensional experiment. It 
should  be pointed out, that, of course, in this scenario electric charge is 
conserved in the full multi-dimensional space \cite{dr, drt1}.
As our experiment is not sensitive to the charge localized outside the brane 
it would detect the charge nonconservation through the $particle \to invisible$ decay. In the example illustarted in Fugure \ref{fig:brane}, a muon 
produced in the $\pi \to \mu + \nu$ decay escapes into extra dimensions
resulting in the $\mu \to invisible $ decay. 
Hence, one may conclude that an observation of the process of a particle
 disappearance 
would provide a strong evidence for the existence of  extra dimensional world.
It may be worthwhile to remember that the process with the 
analogous experimental signature, $Z \to invisible$ decay of the gauge boson $Z$ plays a 
fundamental role in determination of the number of lepton families 
in the SM.

Some examples of charge-nonconserving processes involving leptons and baryons 
 can be found in Particle Data Group \cite{pdg}. Experiments 
 searching for the  $electron \to invisible$ decay  test the electron 
stability  and yield 
lower half-lifetime limits in the range $10^{23}-10^{26}$ yr.   
For more recent results see, also \cite{electron}. One of the most 
stringent limits for charge-nonconserving decays of neutrons, 
$\Gamma(n\to p + \nu_e + \overline{\nu}_e)$/$\Gamma(n\to p + e^- + \overline{\nu}_e)\leq 8 \times 10^{-27}$ has been extracted from the reported counting 
rates of solar neutrino experiments \cite{norman}.  
As far as  the charge-nonconserving $\mu, \tau \to invisible$  
decay modes are concerned, they have never been experimentally tested.

In extra-dimensional models the rate of processes like 
$e, \mu, \tau \to invisible$, is  small and might be  enhanced 
for higher masses, 
  but presently cannot be reliably 
predicted as it is depended on unknown parameters of extra-dimensional 
physics \cite{drt1, rub, dub2}. 
It would be interesting to  perform direct high sensitivity searches for 
these unexpected decay modes, whose discovery would lead to discovery of 
new physics. 

In this  paper we obtain the first limit on the electric 
charge nonconservation in muon and tau  decays and  show that 
the muon bound  can be significantly improved in a  new 
proposed experiment.

\section{Limit on the $\mu^+ \to invisible$ decay}

Recently, the MuLan collaboration has reported on measurements
of the mean lifetime $\tau_\mu$ of positive muons to a precision of 11 ppm
\cite{mulan}. Using 
the new world average 
\begin{equation}
\tau_{\mu} = 2.197 019(21)~ \mu s
\label{taumu}
\end{equation}
and the relation between the muon lifetime and the Fermi constant $G_F$
\begin{equation}
\tau_\mu^{-1}= \frac{G_F^2 m_\mu^5}{192 \pi^3}(1+\Delta)
\label{width}
\end{equation}
where $\Delta$ is the sum of phase space, QED and hadronic corrections, 
results in  new determination of the Fermi constant 
\begin{equation}
G_F = 1.166 371(6) \times 10^{-5}~ GeV^{-2}
\label{gf} 
\end{equation}
to a precision of 5 ppm  \cite{mulan}.

If the $\mu \to invisible$ decay 
exists, it would contribute to the total muon decay rate:
\begin{equation}
\tau_\mu^{-1}= \Gamma_\mu (\mu \to all) = \Gamma_{SM} + \Gamma(\mu \to invisible)+...
\label{rate}
\end{equation} 
and, hence increase the determined value of $G_F$.
To estimate the allowed contribution of $\Gamma(\mu \to invisible)$ to Eq.(\ref{rate}),
one could compare the measured  muon decay rate of Eq.(\ref{taumu})
to a predicted muon decay rate, calculated from Eq.(\ref{width}) 
by using $G_F'$  extracted from another measurements which are 
not affected  by the disappearance effect discussed above.  

 There are a number of indirect prescriptions for extracting of precise values
of $G_F$. Comparison of these quantities can be used to 
to test the SM and to probe for new physics, for more 
detail discussions see Ref. \cite{marciano}. For example, one can define 
\begin{equation}
G_F' = \frac{4 \pi \alpha}{\sqrt{2} m_Z^2 sin^2 2\Theta_W(m_Z)(1-\Delta r)}
\end{equation}
where $\Theta_W$, $m_Z $  and $\Delta r$ are the Weinberg angle, the mass of 
the $Z$ gauge bosons and a factor for radiative corrections, respectively.

Using the values of $\Theta_W$, $m_Z $  and $\Delta r$ reported in  
\cite{marciano}, one can obtain
\begin{equation}
G_F' = 1.1672(\pm 0.0008)\left ( \begin{array}{c}
+0.0018 \\
-0.0007\\
\end{array} \right ) \times 10^{-5}~GeV^{-2}
\label{gf1}
\end{equation}

Comparing  Eq.(\ref{gf}) and Eq.(\ref{gf1}) and adding 
 statistical and systematic errors in quadrature, one finds 
\begin{equation}
\Delta G_F = G_F^\mu - G_F' < 2.6 \times 10^{-3} ~GeV^{-2}~ (90\% C.L.)
\end{equation}
 which  leads to the bound 
\begin{equation}
Br(\mu^+ \rightarrow invisible) = \frac{\Gamma(\mu^+ \to invisible)}{\Gamma(\mu^+ \to all)}< 5.2 \times10^{-3}
\label{muon}
\end{equation}

In close analogy with muon decays, the leptonic decay rates of the tau 
$\Gamma(\tau \to e \nu \overline{\nu})$ and 
$\Gamma(\tau \to \mu \nu \overline{\nu})$
 \cite{pdg} can also provide corresponding Fermi constants \cite{marciano}
\begin{equation}
G_F^{\tau e} = 1.1666(28) \times 10^{-5}~GeV^{-2}
\label{taue}
\end{equation}
\begin{equation}
G_F^{\tau\mu} = 1.1679(28) \times 10^{-5}~GeV^{-2}
\label{taue}
\end{equation}
They   agree with $G_F$ of Eq.(\ref{gf}) within 240 ppm 
and, similar to above considerations,  can be used to constrain the 
$\tau \to invisible$ decay rate. Taking into account  
the branching ratio of the leptonic decay rate of tau, 
the best limit obtained is 
\begin{equation}
Br(\tau \rightarrow invisible) = \frac{\Gamma(\tau \to invisible)}{\Gamma(\tau \to all)}< 1.6 \times10^{-3}
\end{equation}
 
The limit of Eq.(\ref{muon})  could be  
improved by more than seven orders of magnitudes in the proposed experiment 
discussed below, while to improve significantly the bound on  the 
$\tau \to invisible$ decay rate might be  more difficult due to the 
problem with the  efficient tagging of the tau appearance. A special study 
has to be performed in this case.

\section{Direct experimental search for the  $\mu^{+}\to invisible$ decay}

\begin{figure}
\includegraphics[width=0.45\textwidth]{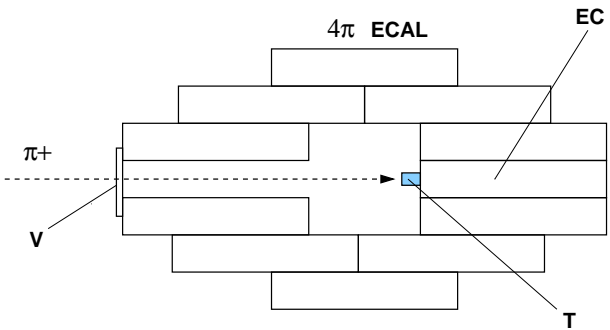}
\caption{\label{fig:setup} Schematic illustration of the experimental setup 
to search for  the $\mu \to invisible$ decay. Shown are the beam defining 
scintillator counter $V$, which is also used as a veto against decay positrons,
the active target $T$, the electromagnetic calorimeter ECAL, and the 
ECAL endcap counter EC used as a light guide for the light produced in the target.}
\end{figure}

The main components of the detector to search  for the 
$\mu \to invisible$ decay are shown in Fig.\ref{fig:setup}.
The basic ideas are as follows. Charged pions are stopped in an active target ($T$)
 instrumented with energy deposition and time readout. 
The source of muons 
is the $\pi \to \mu +\nu$ decay at rest in $T$. 
The target is surrounded by a hermetic 4$\pi$ electromagnetic calorimeter 
(ECAL) to detect energy deposition from the decay $\mu \to e + anything$ of 
muons stopped in $T$.  The light signals produced in  $T$ could be  readout through 
the  ECAL endcap crystal ($EC$) which acts as a light guide  as shown in Fig.\ref{fig:setup}.
The $T$ signals could be distinguished from the $EC$  signals due to their significantly 
different decay times by using the technique described 
in detail in Ref.\cite{bader}. 
  
The readout of the energy deposition in the ECAL 
is triggered by a tag signal of the muon 
appearance, which is defined by a coincidence of a 
 signal from a stopped pion and a delayed signal from the  
 stopped decay muon. The latter should correspond to the 4.2 MeV 
energy deposition in the target from the muon kinetic energy. 

The experimental signature of  the $\mu \rightarrow invisible$ decay
is the apparent disappearance  of the energy deposition in the ECAL from
 the ordinary  electric charge-conserving 
$\mu \to e + anything$ decay, i.e.  observation of  an event with 
the ECAL energy deposition  below of a certain threshold $E_{th}$.   

To estimate the sensitivity of the proposed experiment 
 a feasibility study  based on GEANT4 \cite{geant}
Monte Carlo simulations
have been  performed. However,
 to investigate the background down to the level  $Br(\mu \to invisible) \lesssim 10^{-10}$ with the full simulation would require 
the generation of a very large number of muon decays resulting in a
prohibitively large amount of computer time. Consequently, only the most 
dangerous background processes are considered  
and estimated  with a  smaller statistics combined  with numerical calculations.

The beam of positive pions is 
stopped  in the central part of the cylindrical  target  
in a volume of $\lesssim1$ cm$^3$.
The active target $T$ is a plastic scintillator with 
a diameter of 5-10 mm and a height of 10 mm. According to simulations  the 4.2 MeV muon came to rest passing about 1 mm in $T$.  The  ECAL is an array of $\simeq 100$  BGO counters each of 
52 mm in diameter and  220 mm long, which was  previously used in the PSI experiment on 
precise measurements of the $\pi \to e+\nu $ decay rate \cite{pienu} and then in  
the ETH-INR positronium experiment  \cite{bader}.  The following
 sources  of background are considered
\begin{itemize} 
\item the principal muon decay $\mu \to e \nu \overline{\nu}$
 into the final state 
with the electron kinetic energy $E_{kin}$  less than the 
 detection energy threshold $E_{th}$ (typically 100-300 keV). Indeed, if  
 $E_{kin} < E_{th}$ the event   
 becomes invisible. The partial muon decay rate $\Delta \Gamma_\mu$ 
into the such final state, assuming  $E_{kin} < m_e$ (here, $m_e$ is the electron mass),  is 
\begin{equation}
\Delta \Gamma_\mu \simeq \Bigl(\frac{p_{e}}{p_{e}^{max}}\Bigr)^4 \Gamma_\mu=3.6\times 10^{-4} \Bigl(\frac{E_{th}}{m_\mu}\Bigr)^2 \Gamma_\mu
\end{equation}
where $p_e, p_e^{max}$ is the momentum and the maximum allowed momentum 
of the decay electron, respectively and $ p_e \simeq (2 m_e E_{th})^{1/2}$ . 
 To suppress this background, one has to use  as low as possible threshold 
$E_{th}$ and 
to  performed the experiment with a
well separated {\it positive} pion beam 
with an extremely small contamination of negative  
pions (or muons). In this case, if positive muon 
decays into a low energy positron, the latter will stop in the target and 
annihilate into two photons at a lifetime scale of the order of a few ns.
Thus, for events with $E_{kin}<E_{th}$ 
 the  minimum energy deposition in the ECAL 
will be about 1 MeV, i.e. well above the 
threshold, thus making these  events visible. 
For example, in the ETH-INR experiment  the probability  
to observe $2\gamma$- annihilation energy deposition in the BGO ECAL 
less than the
energy threshold  of $E_{th}\simeq 80$ keV was about 
$ P_{2\gamma} \simeq 10^{-8}$ \cite{bader, pc}.
Thus, in the background free experiment 
 one potentially can reach sensitivity in the
branching ratio of the invisible muon decay  as small as:
\begin{equation}
Br(\mu \to invisible) \sim \frac{\Delta \Gamma_\mu}{\Gamma_\mu} \cdot P_{2\gamma} \simeq 10^{-18}
\label{sens}
\end{equation}
assuming  the detection energy threshold to be as low as 
$E_{th} \simeq 100$ keV.

\item fake tags of the muon appearance could be due to the following effect.
A muon from the pion decay in flight stops in $T$ and decays quickly, within, 
say, few tens of ns into a low-energy positron which subsequently  
came also to rest in $T$ and annihilates into two photons.
If the sum of energies deposited by the positron and  annihilation photons, 
due to their photo-absorption or Compton scattering 
  in $T$,  is around 4.2 MeV this results in the fake muon tag signal.
In the case when the positron kinetic energy 
is about 3 MeV and almost all annihilation energy is deposited in $T$
the event becomes invisible.
To suppress this background the target should be optimized in size and 
made of a low-Z material to minimize 
cross-section of the photo-absorption which is $\sigma_{pha}\sim Z^5$. For example, for plastic 
scintillator the probability of both 511 keV photons energy absorption 
in a volume of $\simeq1$ cm$^3$ is found to be less than 10$^{-8}$.
The size of the target is important in order to minimize the number 
of low-energy positrons stopped in $T$. The Monte Carlo simulations of low-energy positrons 
suggests that for suppression of this background 
 the beam spot size and the diameter of $T$ have to be as small as possible.
If the $T$ diameter
is $\simeq 5$ mm,  the requirements to get simultaneously the 4.2 MeV deposited
 energy and the positron track stopped in $T$  results in  suppression 
of this background processes to the level $\lesssim 10^{-11}$.      

\item  the loss of the muon decay energy 
 due to non-complete hermeticity of the ECAL or due to the presence 
of passive materials. In the present version Monte Carlo simulations
include active materials of the ECAL and the target 
and negligible amount of passive materials from 
the ECAL crystals and the target wrapping. It is found 
that the most dangerous background process is associated with energetic positrons
escaping the ECAL though the beam entrance aperture. To suppress this background 
the aperture should be reduce to as much as possible size and 
should be  closed by a high efficiency  veto counter $V$, as shown in Fig.\ref{fig:setup}, which 
could also act as the beam defining counter.

An additional suppression factor came form the fact that the backward moving 
decay positrons deposit  in $T$  about 1 MeV in addition to 4.2 MeV 
deposited by stopped muons.  Assuming 10\% resolution (FWHM) for measurements 
of energy in $T$, the diameter of the entrance aperture of 1 cm and 
the inefficiency of $V$ $\simeq 10^{-3}$ 
leads to  the final suppression of this source of background down to the level
$\lesssim 10^{-11}$.
\end{itemize}

Finally let us discuss  several additional limitation factors.
The first one is related to the 
relatively long muon lifetime. For example, 
 to get  the branching ratio
$Br(\mu \to invisible) \simeq 10^{-10}$ , the ECAL gate duration  
$\tau_g$, and hence the dead-time per trigger, has to be 
\begin{equation} 
\tau_g \gtrsim - \tau_\mu \times ln(Br(\mu \to invisible)) \simeq 50 ~ \mu s
\label{dur}
\end{equation}
 in order to avoid 
background from the muon decays outside the gate. 
In the ETH-INR positronium experiment, the ECAL gate $\tau_{Ps}$ 
was about $\simeq 2~\mu s$ for orthopositronium lifetime in the target of 132 ns. This 
resulted in distribution of the sum of pedestals of all ($\simeq 100$) 
ECAL counters corresponded to the threshold of 
80  keV used to define the signal range for the 
$o-Ps \to invisible$ decay. In the proposed  experiment the
 longer gate will lead to 
 an increase of the pile-up and pick-up electronic 
noise and hence to the overall
broadening of the  signal range,  approximately by a factor  
$\sqrt{\tau_g/\tau_{Ps}}\simeq 5$
and, hence to an increase 
of the energy threshold roughly up to $E_{th}\simeq 400$ keV  \cite{pc}. 
For this threshold the probability of the annihilation energy loss is 
about $P_{2\gamma} \simeq 10^{-4}$ \cite{pc} and the overall sensitivity 
of Eq.(\ref{sens}) drops to a few parts in $10^{13}$.

Another limitation factor is related to the dead time of  Eq.(\ref{dur})
and, hence to  the maximally allowed muon 
counting rate, which  according to  Eq.(\ref{dur})  
 has to be  $ \lesssim 1/ \tau_g \simeq 10^{4} \mu / s$ 
to avoid significant pile-up effect.    
To avoid this limitation,   one could 
 implement a fast first-level trigger rejecting events 
with the ECAL energy deposition greater than $E_{th}$ and, hence  run the 
experiment at the rate $\simeq 1/\tau_\mu \simeq  10^5~\mu/s$.
 Thus, in the background free experiment 
 one could expect a sensitivity in the $\mu \to invisible$
 decay branching ratio of the order of $10^{-11}$, assuming 
that the  exposure to the muon beam with this rate is  $\simeq$ 1 month. 
The performed Monte Carlo simulations 
 give an illustrative correct order of magnitude for the sensitivity of the 
proposed experiment
and may be strengthened  by more accurate and  detailed Monte Carlo 
simulations of the concrete  experimental setup.\\

\section {Conclusion}
In this work the  first limits on the electric charge-nonconserving 
$\mu, \tau \to invisible$
decay modes are obtained. If the   $\mu^+ \to invisible$  decay exists
at the level  of a few parts in 10$^{11}$, it could be observed 
in the new proposed experiment. The preliminary study shows that the
quoted  sensitivity could be obtained with a setup optimized for  
several its properties. Namely, i) the primary beam and the entrance aperture 
size, ii) the energy resolution, material composition 
and dimensions of the target, iii) the
efficiency of the veto counter, and iv) the pile-up effect and zero-energy 
threshold in the ECAL are of importance.
 This low-energy experiment might be a sensitive  probe of extra-dimensional
 physics that is complementary to  collider experiments. 

{\large \bf Acknowledgments}

The author is grateful to S.L. Dubovsky for stimulating discussions, M.M. Kirsanov for help in simulations and N.V. Krasnikov for encouragement which made this 
article possible.
It is a pleasure to acknowledge  P. Crivelli, 
N.V. Krasnikov,  V.A. Matveev, V.A.  Rubakov, A. Rubbia, D. Sillou and F. Vannucci for useful comments.

\end{document}